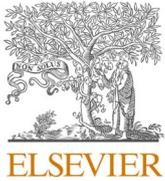
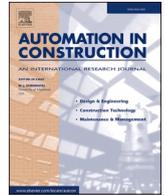

# Multi-system intervention optimization for interdependent infrastructure

Omar Kammouh, Maria Nogal[*], Ruud Binnekamp, A.R.M. Rogier Wolfert

*Dept. of Civil Engineering and Geosciences (CEG), Delft University of Technology, Delft, the Netherlands*

ABSTRACT

The wellbeing of modern societies is dependent upon the functioning of their infrastructure networks. This paper introduces the 3C concept, an integrative multi-system and multi-stakeholder optimization approach for managing infrastructure interventions (e.g., maintenance, renovation, etc.). The proposed approach takes advantage of the benefits achieved by grouping (i.e., optimizing) intervention activities. Intervention optimization leads to substantial savings on both direct intervention costs (operator) and indirect unavailability costs (society) by reducing the number of system interruptions. The proposed optimization approach is formalized into a structured mathematical model that can account for the interactions between multiple infrastructure networks and the impact on multiple stakeholders (e.g., society and infrastructure operators), and it can accommodate different types of intervention, such as maintenance, removal, and upgrading. The different types of interdependencies, within and across infrastructures, are modeled using a proposed interaction matrix (IM). The IM allows integrating the interventions of different infrastructure networks whose interventions are normally planned independently. Moreover, the introduced 3C concept accounts for central interventions, which are those that must occur at a pre-established time moment, where neither delay nor advance is permitted. To demonstrate the applicability of the proposed approach, an illustrative example of a multi-system and multi-actor intervention planning is introduced. Results show a substantial reduction in the operator and societal costs. In addition, the optimal intervention program obtained in the analysis shows no predictable patterns, which indicates it is a useful managerial decision support tool.

**Notations**

$c_k \in \mathbb{R}^+$ cost of performing an intervention of type $k$.
$cl_i \in \mathbb{R}^+$ service unavailability cost of object $i$
$f_1 \in \mathbb{R}^+$ total cost of interventions
$f_2 \in \mathbb{R}^+$ total service unavailability cost caused by the interventions
$G_{\min, k} \in \mathbb{N}^+$ minimum number of time steps between two successive interventions of intervention type $k$
$G_{\max, k} \in \mathbb{N}^+$ maximum number of time steps between two successive interventions of intervention type $k$
$\mathbf{I} = [I_{ij}]$ matrix of interaction between objects $i$ and $j$.
$K \in \mathbb{N}^+$ number of intervention types
$\mathbf{M} = [m_{k, t}]$ matrix indicating the existence of intervention of type $k$ at time step $t$
$n_I \in \mathbb{N}^+$ number of intervention types
$N \in \mathbb{N}^+$ number of analyzed objects (e.g., water pipe, road section, etc.)
$\mathbf{R} = [r_{ik}]$ relation matrix indicating upon which object $i$ each intervention type $k$ intervenes
$T \in \mathbb{N}^+$ number of time steps considered
$\delta(.)$ Kronecker delta

## 1. Introduction

Infrastructure networks are subject to constant degradation due to their excessive use and natural hazards [1,2]. Degradation of infrastructure networks ultimately leads to failure, which affects the service quality and causes safety issues and physical damage. Interventions such as maintenance and renovations are executed ensuring a continuous fulfillment of the infrastructure functional goals and its related quality of service parameters (e.g., water protection, traffic flow, etc.). Infrastructure managers are increasingly realizing that the availability of their infrastructures can be effectively increased by better planning their interventions [3]. It is being realized that budget availability is not the only important factor when planning effective intervention programs that aim at increasing infrastructure availability, optimal planning is as important and could yield a high infrastructure availability with a reduced budget. Therefore, when dealing with infrastructure intervention planning, another way of thinking is required.

*1.1. System thinking*

Current approaches dealing with infrastructure intervention planning employ *analytical thinking* rather than *system thinking*, which is considered a more appropriate approach. In analytical thinking, the object to be studied is treated as "a whole to be taken apart", while in system thinking, the thing to be analyzed is treated as "a part of the containing whole"; the first reduces the focus while the second expands it [4]. In system thinking, the performance of a system depends more on how its parts interact than on how they act independently of each other [2]. In the context of infrastructure intervention planning, the intervention program of an infrastructure network is normally obtained by conducting a deterioration analysis for each object in the network [5–7]. The results of the deterioration analysis lead to optimal intervention programs for the individual objects. While these results guarantee a low risk of failure for the individual objects, they might not be optimal when considering the system as a whole, causing frequent system disruptions. Following system thinking, this can be resolved by grouping interventions, even if this results in sub-optimal intervention programs for the individual objects. The benefit is mainly due to the reduced service interruptions and reduced intervention costs.

Literature offers state-of-the-art approaches on how to combine interventions to reduce the total cost incurred by the infrastructure operator. These approaches can be classified into system-specific approaches that focus on specific types of engineering systems, such as bridges [8], roads [9], pavement [10], water networks [11], and offshore wind turbines [12]; and general approaches that can be applied to different (single) engineering systems [13–15]. Most of these approaches use optimization models that take advantage of the set-up and crew travel costs to reduce the total cost. The aim is to reduce the financial impact on the owner by decreasing the cost of performing a set of interventions. Therefore, they do not consider other types of impact that can result from executing interventions, such as user impact and impact on objects that are of spatial proximity.

*1.2. Service unavailability impact*

The unavailability of an object in an infrastructure network, due to either intervention or unexpected disruption, causes an impact on different stakeholders. Many attempts to consider the service unavailability impact on multiple stakeholders in the planning process fail to do so. In the context of this paper, a stakeholder is defined as an individual, group, or organization that is directly or indirectly connected to the infrastructure network Adey, et al. [16]. A person driving a vehicle on a road would be considered as a user that is directly affected by the unavailability of the road. The same person when is far from the road would be considered a part of the indirectly affected public.

Adey, et al. [16] and Kerwin and Adey [17] distinguish between different types of service disruption impact for different stakeholders. For example, comfort, noise, and accidents are users' impact types, while the cost of intervention is an owner's impact type. While the focus of their papers is the service of road and water networks, respectively, most of the ideas presented hold for other types of infrastructure networks. Hence, this paper adopts the definition and classification of service unavailability impact in [16]. In addition, an often-used practice when planning intervention programs for an infrastructure network is to consider the impact on the owner in terms of intervention cost [16]. Considering only the intervention cost implicitly implies that none of the other types of impact matters. To allow for a net benefit maximization, all impact types resulting from service interruption must be considered [18,19].

Quantifying the value of lost service, which depends on how different stakeholders value the services as well as how long and in which way the services are interrupted, represents a challenging task. Kielhauser, et al. [20] proposed a methodology to estimate the loss of service of infrastructure networks due to deterioration and intervention.

Other authors have proposed mathematical equations to quantitatively estimate the impact of service loss on multiple stakeholders, such as [17] for water networks and [16] for road networks.

*1.3. Spatial proximity and interconnectivity of infrastructure networks*

As infrastructures have become gradually interconnected, traditional ways of maintaining them turned out to be insufficient. It has become increasingly evident that infrastructure interconnectivity plays a large role. The unavailability impact of infrastructure services due to intervention becomes apparent when the intervention plans of multiple operators do not go in harmony with one another. What if, for instance, executing an intervention on the water pipe necessitates closing a road for excavation when no intervention is planned for the road itself? This implies users suffering the road unavailability twice, the first is when the water pipe is replaced and the second is when intervention is scheduled for the road. To increase the global system operability, the road intervention can be scheduled at the same time the intervention on the water pipe is scheduled. While in this case the road operator would have to pay extra cost due to the early intervention, the net benefit can significantly increase. Therefore, intervention programs for spatially-close and functionally-connected networks should be developed together. Coordinating infrastructure intervention is progressively becoming of paramount importance to reduce service disruption and repair costs.

Most of the available intervention planning approaches target individual infrastructures [21–26]. These studies focus on grouping interventions related to a single infrastructure type without considering the interconnectivity with other infrastructure types that are on spatial proximity. There has been little research on the determination of optimal intervention programs for interrelated infrastructure networks, i.e. where events on one network affect other networks [18,20,27–29]. For example, Kielhauser, et al. [27] proposed two methods to account for infrastructure interconnectivity when planning an intervention program. Both methods allow for consideration of the spatial proximity of the objects within the infrastructure networks. Kielhauser and Adey [29] presented a unified model of the service provided by infrastructure networks to be used in the search for optimal intervention planning. Kielhauser, et al. [20] developed a methodology to estimate the loss of service on several infrastructure networks due to both preventive and corrective interventions. The aforementioned studies focus more on infrastructure interconnectivity due to spatial proximity (geographical) than on other types of interactions (e.g., functional, cyber, logical, etc.) These studies are also limited to direct interconnectivities of infrastructure objects. The case when an intervention on an infrastructure object (e.g., water pipe) indirectly affects another infrastructure object (e.g., road section) is not covered. This can occur when an intervention on a water pipe requires closure of given road section (road section A), which is in series with another road section (road section B). This implies that an intervention on the water pipes directly affects road section A and indirectly affects road section B. Capturing this progressive service unavailability impact among infrastructure networks represents a current challenge.

*1.4. Modeling error*

Another aspect the existing research does not take into account is the centralized nature of some intervention types. Some of the current intervention planning models are based on an unrealistic assumption that all interventions are flexible to be moved forward or backward in time when clustering interventions [30,31]. This is rather too optimistic because there are interventions that can only be implemented at certain instances in time and therefore they cannot be moved. Take for example the intervention scheduling of railway tracks under limited possession time granted by the infrastructure manager. In this case, the possession-based intervention is usually 'given'; i.e., imposed as a single option, and scheduled well in advance and cannot be moved forward or backward.

Another example is the closure of highway tunnels which is often primarily driven by the time- or usage-dependent cleaning which takes place at fixed intervals. Accounting for such intervention types, which are called hereafter *central* interventions, can be done by fixing those interventions at predefined moments through a process called *centralizing*. All other interventions can then be clustered around the central interventions. Several existing models focus solely on the *clustering* (or grouping) part of the problem when *calculating* the optimal intervention plan, ignoring the *centralization* part (see for instance [14]).

*1.5. Research goals & novelties*

The primary goal of this paper is to cover all the previously mentioned shortcomings of existing scientific literature by introducing an integrative multi-system optimization approach for infrastructure interventions in which multiple infrastructure networks and stakeholders can be reflected. To summarize, this paper:

1) Formalizes a mathematical model and simulation approach for intervention scheduling from a system-level perspective to be able to objectify the most effective and efficient intervention programs;

2) Considers multiple infrastructure networks and accounts for the hindrance impact caused by the interventions, accommodating different types of interventions, such as maintenance, upgrading, and removal;

3) Accounts for central interventions that are usually implemented with a fixed time interval, reflecting the real-life human intervention planning actions and systems behavior;

4) Models the interdependency within and across infrastructure networks by introducing the interaction matrix (IM);

5) Illustrates the applicability of the introduced approach using a simple multi-system intervention plan composed of three interacting networks (i.e., highway, railway, and water network).

The aforementioned approach is called the integrative 3C concept, where '3C' covers three main elements in the proposed intervention planning: *centralize*, *cluster*, and *calculate*. The 3C concept builds upon recent operational research in the field of intervention scheduling and optimization. To our knowledge, this concept has not been implemented in any of the existing state-of-the-art engineering asset management tools (e.g., IBM Maximo, Oracle, eAM, etc.) or any scientific literature reference in which the set of novel points offered in this paper is covered.

The remainder of the paper is organized as follows. Section 2 presents the 3C concept for optimizing interventions considering the interaction among infrastructure networks. Section 3 introduces the mathematical formulation of the multi-system optimization problem. Section 4 presents a numerical example to illustrate the applicability of the proposed optimization model. Finally, conclusions are drawn in Section 5 together with the proposed future work.

**2. The integrative 3C concept: Centralize, Cluster, Calculate**

In this paper, an integrative systems-thinking approach is followed. This implies that the focus is not placed on the individual objects within an infrastructure network as this could produce plans that are tailored for the objects. The attention is rather placed on the integrated system, which is in the context of this paper the interconnected infrastructure networks. As will be demonstrated later, this vision expansion will give a wider view of the problem by considering the joint benefits of the infrastructure operators (i.e., intervention cost) and the users (i.e., societal impact, hindrance, possession, etc.).

*2.1. 3C concept definition*

This section presents the 3C concept for effectively scheduling interventions of interdependent infrastructure networks. What stands out in this concept is its ability to reflect the actual human actions during intervention planning. This occurs in the first phase of the 3C concept (i. e., *centralization*) where the intervention types are classified into *central* and *non-central*. Central intervention types are those that must occur at a pre-established time moment, where neither delay nor advance is permitted. They are usually implemented with a fixed time interval[1] due to their dominant time-dependent nature. This time interval represents the time between two interventions of the same type; for instance, the time between an intervention on a road section and the next intervention on the same road section. The non-central intervention types are condition-based interventions and can be scheduled during the planned closures of the central interventions. That does not mean anymore defining optimal intervention moments for each intervention type separately, but that the planned interventions (i.e., central interventions) serve as a starting point for the non-central interventions.

The second phase of the 3C concept is the *clustering* phase. In the clustering phase, the non-central interventions are clustered with the planned central interventions while respecting some predefined individual constraints, such as the time interval between two successive interventions of the same type. The first two steps of the 3C concept are illustrated in Fig. 1, where intervention type A is *central* and intervention type B is *non-central*.

Every intervention type is assigned two values, $G_{min, k}$ and $G_{max, k}$, which are two externally imposed constraints that represent the minimum and maximum time intervals, respectively, between two interventions of the same type. $G_{min, k}$ is set to prevent frequent interventions on the same object in the network. It can simply be set to '1' if no minimum requirements are available. $G_{max, k}$, on the other hand, is set to reduce the risk of failure of an object. The risk of failure of an object increases as the time from the last intervention increases [32,33]. $G_{max, k}$ is equivalent to the Mean Time Between Failures (MTBF) of an object. It can be determined using existing methods, such as block replacement models [34], delay-time models [35], and degradation models [6,36,37]. The use of these models is normally limited by the availability of historical data on the average time between failures. If these data are unavailable, the MTBF can be estimated using expert judgment techniques and expert knowledge. Since central intervention types are implemented with a fixed time interval, the minimum and maximum time intervals between two successive interventions of the same type, $G_{min, k}$ and $G_{max, k}$, are set equal. Fig. 2 illustrates the notion of *time interval* for two intervention types A and B, where A is a central intervention type and B is a non-central intervention type.

The third and final phase is to *calculate* or optimize for the inter-vention program that meets the conditions initially set. In this work, the optimization objective is to reduce the global cost of executing the interventions. The global cost is divided into two parts: the first part is the actual cost of the interventions while the second part is the cost of suspending all affected objects (i.e., service unavailability cost). The unavailability cost represents the impact of service unavailability. Implementing multiple interventions at the same time signifies reduced service unavailability, and thus a reduced cost. The unavailability cost of the service must be monetized to be able to combine it with the intervention cost. Several comprehensive methods on how to monetize the impact of service unavailability have been recently published [16,17,38]. The mathematical formulation of the 3C concept is introduced in Section 3.

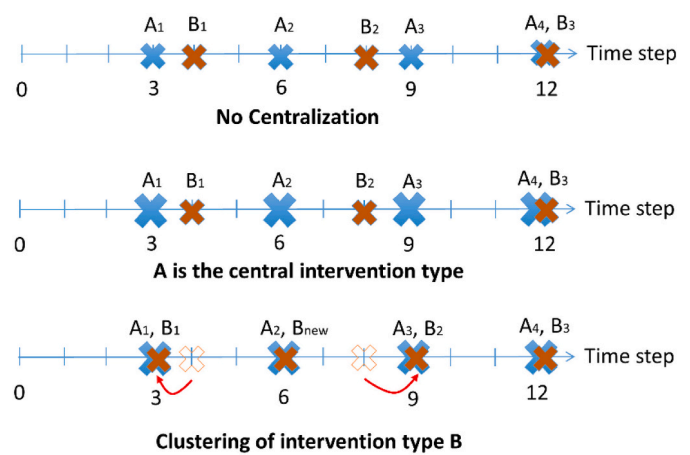

**Fig. 1.** Centralizing and Clustering of interventions.

---

[1] As indicated in the introduction, this time interval can be originated by a given slot of an infrastructure manager or by usage/time-dependent intervention.

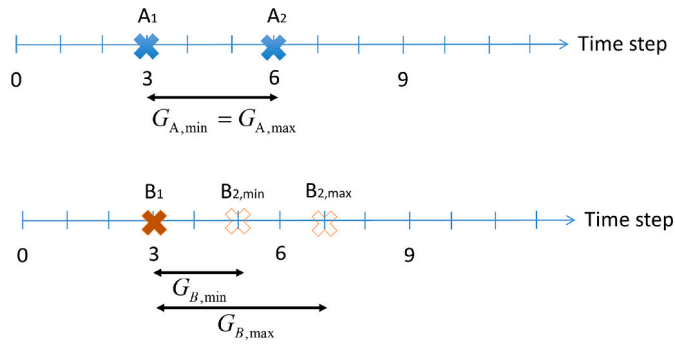

**Fig. 2.** Time interval between two successive interventions of the same type: A is a central intervention type and B is a non-central intervention type.

### 2.2. Multi-system consideration

In the context of this paper, an object (*obj* in Fig. 3) is considered a part of a major infrastructure network (*Net*). The definition of the object will respond to modeling needs in a manner that two contiguous network sections or components characterized by different essential features represent two different objects. An operator is the manager of an infrastructure network, who can be responsible for one or multiple networks. Each operator is assumed responsible for the intervention and service unavailability costs of their networks. An intervention type (*Int*) is an intervention on one or more objects. Fig. 3 shows the relationships between operators, infrastructure networks, assets, objects, and intervention types. The proposed 3C concept takes into consideration the interaction among different objects. This is especially important when tackling interconnected infrastructure networks run by multiple operators. For a single network, there is always dependency between the performance of the network and its objects (i.e., executing an intervention on a major object might require a temporary suspension of part of or the whole network). When tackling multiple networks, the loss of performance of an object within Network A could or could not affect an object within Network B. Therefore, the relationships among the objects across the networks should be considered. The interdependency can be split into physical, geographical, and functional interdependency [39]. Physical interdependency arises from a physical linkage between the input and output of two objects (e.g., two water pipes in series). Geographical interdependency occurs if a local event, such as an intervention, can create state changes of multiples objects. This happens when objects of multiple infrastructures are in spatial proximity (e.g., intervention on a buried pipe requires excavation of the road on top). Finally, functional interdependency occurs if the state of an object depends on the state of another object via a mechanism that is not physical or geographic (e.g., electrical power and railway).

Eq. (1) is an IM of a set of $N$ objects, where $I$ is a square matrix whose components, the so-called interaction coefficients $I_{ij} = \{0,1\}$, determine if object $i$ interacts with object $j$. $I_{ij} = 0$ means that object $i$ does not affect the functionality of object $j$, whereas $I_{ij} = 1$ implies the contrary. Consequently, the diagonal terms of IM are $I_{ii} = 1$ and $I$ can be asymmetric as a result of the non-reciprocal interaction behavior between the objects. The values of the interaction coefficients can be obtained from expert judgment [40,41].

$$\mathbf{I} = [I_{ij}] = \begin{bmatrix} I_{11} & \cdots & I_{1N} \\ \vdots & \ddots & \vdots \\ I_{N1} & \cdots & I_{NN} \end{bmatrix} \quad (1)$$

Let's assume an example of a small network involving a road, a railway, and a buried water pipe (Fig. 4). In the area of interest, the section of the road and the section of the railway are located parallel to each other while the section of the water pipe intersects with the road and the railway (Fig. 4-a). This means that executing an intervention on the water pipe would interrupt the functionality of both road and railway because excavation is needed to access the water pipe. On the other hand, executing an intervention on either the road or the railway does not affect the functionality of the other objects. These interactions between the objects can be mathematically represented using Eq. (2).

$$\mathbf{I} = \begin{bmatrix} I_{11} & I_{12} & I_{13} \\ I_{21} & I_{22} & I_{23} \\ I_{31} & I_{32} & I_{33} \end{bmatrix} = \begin{bmatrix} 1 & 0 & 0 \\ 0 & 1 & 0 \\ 1 & 1 & 1 \end{bmatrix} \quad (2)$$

One could argue that executing an intervention on the water pipe in the section buried under the road does not necessitate suspending the operation of the railway. To tackle this, the water pipe can be treated as two objects (Fig. 4-b). In this case, the water pipe is modeled using multiple objects that interact with other objects according to their physical location and functional dependence. The interaction matrix can then be rewritten as follows:

$$\mathbf{I} = \begin{bmatrix} I_{11} & I_{12} & I_{13} & I_{14} \\ I_{21} & I_{22} & I_{23} & I_{24} \\ I_{31} & I_{32} & I_{33} & I_{34} \\ I_{41} & I_{42} & I_{43} & I_{44} \end{bmatrix} = \begin{bmatrix} 1 & 0 & 0 & 0 \\ 0 & 1 & 0 & 0 \\ 0 & 1 & 1 & 1 \\ 1 & 0 & 1 & 1 \end{bmatrix} \quad (3)$$

It is important to note the functional dependence between water pipes $W_1$ and $W_2$. Hence, the interaction coefficients $I_{34}$ and $I_{43}$ are set 1 because it is assumed that the disruption of a pipe section affects the functionality of the other pipe section.

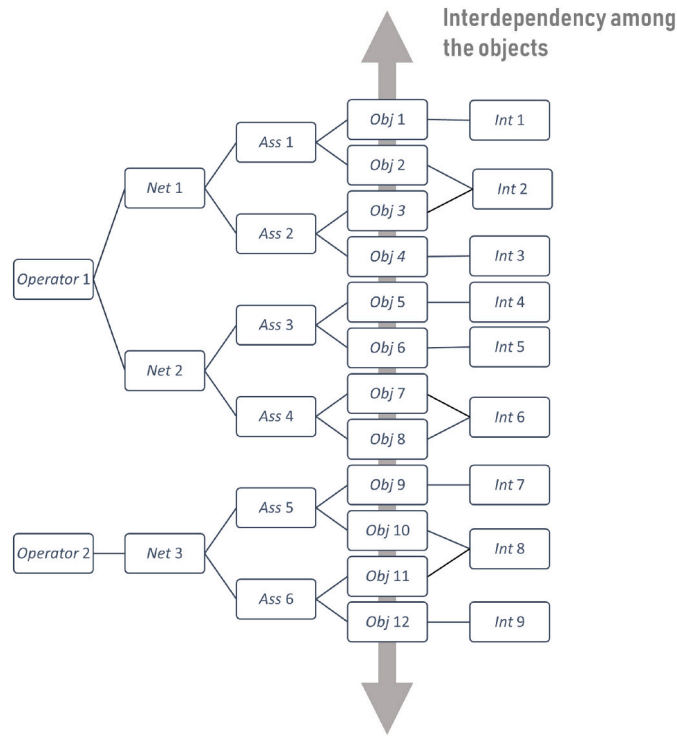

**Fig. 3.** Relationships between operators, infrastructure networks (*Net*), Assets (*Ass*), objects (*Obj*), and intervention types (*Int*).

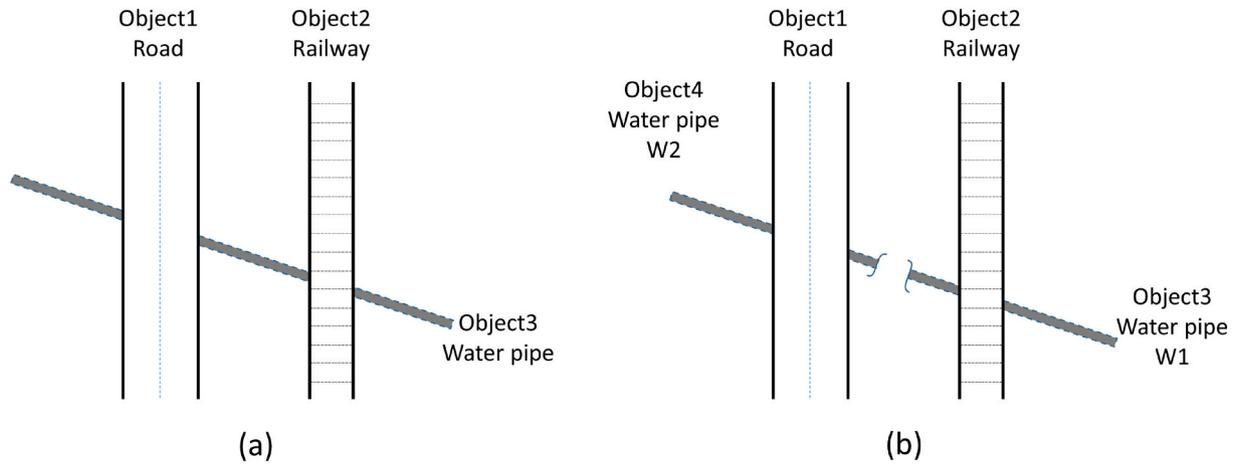

**Fig. 4.** Multi-system interactions (a) water pipe considered as a unique object, (b) water pipe made of two objects.

The matrix *I* can consider any type of interdependency. For example, excavating a road to do an intervention on a water pipe is considered a geographical interdependency. Intervention on a pipe section disrupts the second pipe section on the same pipeline, and this is considered a functional interdependency. Note that the interdependency between objects in a network is not necessarily binary. Intervention on one object can partially affect other objects in the network. In such a case, a value $I_{ij}$ that is between 0 and 1 can be set; however, in this case, Eq. (6) should be modified to account for partial service unavailability costs.

### 3. The mathematical formulation of the 3C problem

#### 3.1. Objective function

The optimization problem aims at scheduling interventions for each object such that the global intervention cost, including the direct intervention costs and the compound costs of service unavailability, is minimum. The optimization problem can be expressed as:

$$\underset{M}{Min} \ (f_1 + f_2), \tag{4}$$

where $f_1$ is the total cost of interventions, given by:

$$f_1 = \sum_{t=1}^{T} c_k \mathbf{M} = \sum_{t=1}^{T} c_k [m_{k,t}] \tag{5}$$

where $c_k \in \mathbb{R}^+$ is the cost of performing an intervention of type $k$ with $k = 1, 2, \ldots K$, $K \in \mathbb{N}^+$ being the number of intervention types, $T \in \mathbb{N}^+$ is the number of time steps considered in the analysis, and $m_{k,t} \in \{0,1\}$ are the components of **M** indicating at which time steps each intervention type is conducted over the total time of analysis. It is assumed that each intervention is entirely performed within a time interval; and $f_2$ is the total service unavailability cost caused by the interventions:

$$f_2 = \sum_{t=1}^{T} cl_i \delta(\mathbf{I}^t \times \mathbf{R} \times \mathbf{M}) = \sum_{t=1}^{T} cl_i \delta\left([I_{ij}]^t [r_{i,k}] [m_{k,t}]\right) \tag{6}$$

where $cl_i \in \mathbb{R}^+$ is the service unavailability cost of object $i$, $[I_{ij}]^t$ is the transpose of $[I_{ij}]$, $r_{i,k} \in \{0,1\}$ are the components of the relation matrix **R** that indicates upon which object $i$ each intervention type $k$ intervenes. The function $\delta(.)$ represents the Kronecker delta defined as follows

$$\delta(x) = \begin{cases} 0 \ if \ x = 0 \\ 1 \ if \ x \neq 0 \end{cases} \tag{7}$$

Note that the Kronecker delta is applied to each component of the resulting ($N$ x $T$) matrix. The Kronecker delta in Eq. (7) allows the consideration of the benefits of clustering interventions, as it introduces the service unavailability cost of an affected object only once when several interventions affecting its performance are occurring at the same time. It is important to note that the Kronecker delta does not cause double-counting for two objects which are in a row. Each object is associated with an independent unavailability cost. That is, if a pipe with an unavailability cost x is divided into two equal sections, each one will be assigned an unavailability cost equal to x/2. The disruption of any of the pipe sections will automatically affect the second pipe section (see the relation matrix **R** in Eq.(6)), and this causes the unavailability cost to be again equal to x. The pipe division serves only the geographic interdependency, for example when a road is crossing only a part of the pipeline. In this case, it is important to identify which part of the pipe is *directly* disrupted by an intervention to know whether the road will also be disrupted.

#### 3.2. Constraints

The first constraint set, which restricts any two successive interventions of type $k$ to have at least a time interval equal to $G_{min, \, k}$, is expressed by Eq. (9), where $G_{\min, \, k} \in \mathbb{N}^+$ is the minimum number of time steps between two successive interventions of intervention type $k$:

$$0 \leq \sum_{t}^{t+G_{min,k}-1} m_{k,t} \leq 1 \qquad for \ t = 1 \rightarrow T - G_{min,k} + 1, \quad \forall k = 1, 2, \ldots, K \tag{8}$$

It is assumed that each intervention is performed and finalized within a time interval. The second constraint set restricts any two successive interventions of type $k$ to have a time interval not larger than $G_{max, \, k}$, as shown by Eq. (8), where $G_{\max, \, k} \in \mathbb{N}^+$ is the maximum number of time steps between two successive interventions of intervention type $k$:

$$\sum_{t}^{t+G_{max,k}-1} m_{k,t} \geq 1 \qquad for \ t = 1 \rightarrow T - G_{max,k} + 1, \quad \forall k = 1, 2, \ldots, K \tag{9}$$

#### 3.3. Optimality and uniqueness of the solution

The mathematical problem defined by Eqs. (4) to (9) is a mixed-integer nonlinear optimization problem where the variables ($m_{k, \, t}$) are binary. This problem can be solved using exact methods, e.g., the Mixed-Integer Nonlinear Programming (MINP). However, exact methods are suitable for problems with a small number of variables. For problems with a medium to a large number of variables, other heuristic approaches can be used. Genetic Algorithm (More precisely, the Integer Genetic Algorithm (IGA)) has been chosen among other (Meta-)heuristic algorithms because of implementation reasons. This algorithm is commonly present in many of the commercial software (e.g., Matlab and Python). The difference between IGA and the ordinary Genetic Algorithm (GA) is the special creation, crossover, and mutation functions of the IGA that enforce the variables to be integers [42]. IGA attempts to minimize the penalty function, not the fitness function. The penalty function contains a term for infeasibility. This penalty function is combined with a binary tournament selection to select individuals for subsequent generations [43]. In this work, the optimization problem can be of a high number of variables. Therefore, IGA is used for solving the optimization problem.

IGA is a heuristic approach that yields the so-called "best-known solution", which is not necessarily an optimum but a near-optimum solution. Given that the IGA is sensitive to the initial population used to seed the genetic algorithm, the optimization problem is solved multiple times with different feasible initial populations and then the (near) optimal solution is chosen among the solutions obtained in each iteration. In this way, the quality of the solution is improved. Optimization based on (Meta-)heuristics always requires a trade-off between the computational time and the quality of the solution. The robustness of the solution must therefore be carefully investigated.

The system of Eqs. (4) to (9) provides a solution space rather than a unique solution, that is, multiple configurations (i.e., intervention programs) can yield the minimum cost. Given that the presented 3C mathematical framework aims to help operators define their best strategies, multiple (near) optimal solutions might be more interesting for the operators. Nonetheless, further restrictions can be imposed to reduce the solution space, and eventually obtain a unique solution; for instance; restrictions regarding the temporal order of performing the interventions, or deadlines to perform some interventions.

## 4. Demonstrative example of the 3C concept

The purpose of this section is to demonstrate the applicability of the 3C concept for preventive intervention planning based on a real-life example from industry. This example has been translated for demonstrative purposes. The example considers different infrastructure networks comprising water, railway, and highway objects. The input data used in this example are realistic in the sense of the proportionality of used values, and therefore they reflect reality and serve well for the objective of this example. Moreover, some of the data, such as the intervention/deterioration data, are relative rather than absolute (e.g., time, euros, etc.).

### 4.1. Networks description and interaction

The network is composed of three infrastructure networks: water network, highway, and railway (Fig. 5). The networks are operated by independent operators, herein called W (water operator), H (Highway operator), and R (Railway operator). Every network is divided into several objects identified according to their physical location and/or functional dependence. The water network is divided into six objects (W1–6), the highway network is divided into four objects (H1–4), and the railway network is divided into two objects (R1–2).

Fig. 5 shows the analyzed infrastructure networks with several intervention types that are to be planned. As shown in the figure, the objects intersect at different locations. These intersections imply interdependency among the objects in such a way that an intervention on one object could cause a closure to the intersecting objects.

Table 1
Data of the analyzed objects.

| Object | Index ($i$) | Service unavailability cost of object $i$, $cl_i$ (monetary unit) $\times 10^3$ | Interaction with other objects ($i$) | Start joint | End joint |
| --- | --- | --- | --- | --- | --- |
| W1 | 1 | 25 | 2 (W2), 7 (H1) | J4 | J5 |
| W2 | 2 | 12,5 | 1 (W1), 11 (R1) | J5 | J6 |
| W3 | 3 | 20 | 9 (H3) | J6 | J8 |
| W4 | 4 | 22 | 10 (H4) | J8 | J13 |
| W5 | 5 | 15 | 6 (W6), 10 (H4), 12 (R2) | J9 | J10 |
| W6 | 6 | 27,5 | 5 (W5), 8 (H2) | J10 | J11 |
| H1 | 7 | 15 | – | J1 | J2 |
| H2 | 8 | 25 | – | J2 | J3 |
| H3 | 9 | 12,5 | 11 (R1) | J2 | J7 |
| H4 | 10 | 20 | 12 (R2) | J2 | J12 |
| R1 | 11 | 22,5 | 9 (H3), 12 (R2) | J13 | J15 |
| R2 | 12 | 15 | 10 (H4), 11 (R1) | J15 | J16 |

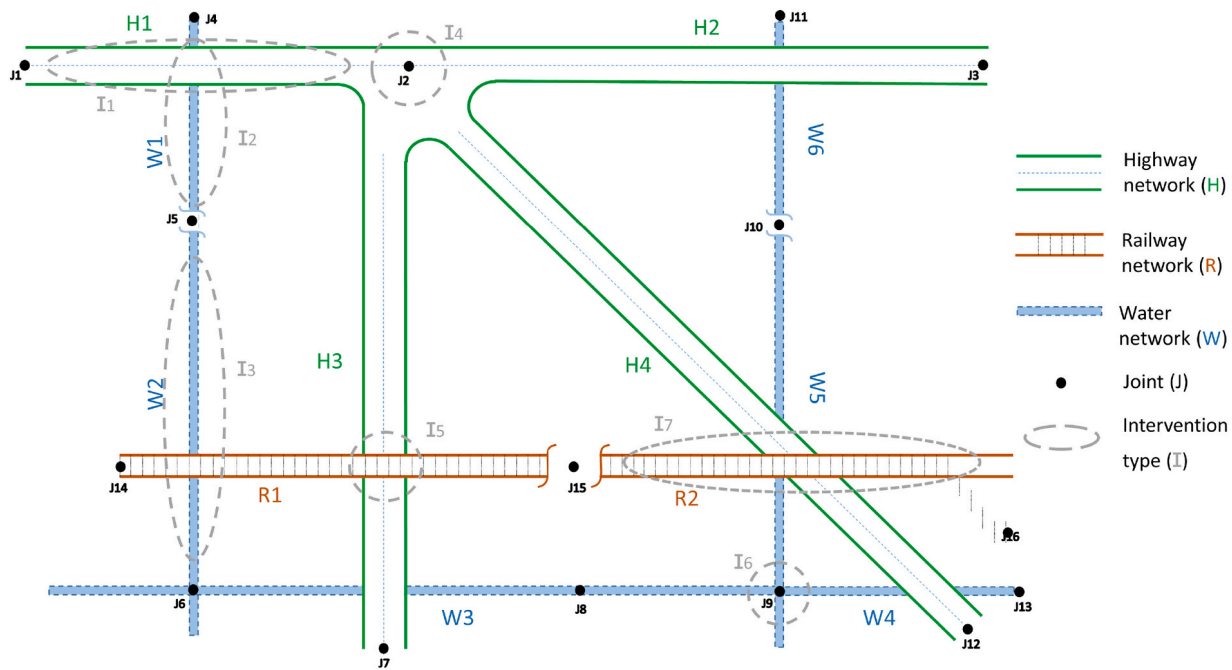

**Fig. 5.** Infrastructure networks with preventive interventions to be planned.

The unavailability of an object incurs service unavailability cost. Table 1 lists the analyzed objects with their service unavailability cost, $cl_i$. Both intervention and service unavailability costs have been proposed by the authors. Service unavailability costs are assumed to be significantly larger than the intervention cost, based on [44–46]. The unavailability cost occurs every time an object is directly or indirectly affected by one of the interventions. Direct effect means that an intervention is executed directly on the object while indirect effect means that an intervention is performed on an interdependent object causing closure to the analyzed object. The service unavailability cost of an object reflects the impact the object would cause if it becomes unavailable (i.e., higher cost implies increased impact). These interactions among the objects, which are necessary to feed the interaction matrix $I$, are also listed in the table. An example of directly and indirectly affected objects is intervention I2 (see Fig. 5). I2 is an intervention type performed on the buried water network object W1, which requires the closure of highway H1 because of the excavation, and the closure W2 because it is an extension object of W1. In this case, W1 is said to be directly affected by I2 while H1 and W2 are indirectly affected. The service unavailability costs of W1, W2, and H1 are thus applied. Eq. (10) presents the interaction matrix $I$, which is derived from Table 1 following Section 2.2.

$$\mathbf{I} = [I_{ij}]$$

$$= \begin{bmatrix}
 & W1 & W2 & W3 & W4 & W5 & W6 & H1 & H2 & H3 & H4 & R1 & R2 \\
W1 & 1 & 1 & 0 & 0 & 0 & 0 & 1 & 0 & 0 & 0 & 0 & 0 \\
W2 & 1 & 1 & 0 & 0 & 0 & 0 & 0 & 0 & 0 & 0 & 1 & 0 \\
W3 & 0 & 0 & 1 & 0 & 0 & 0 & 0 & 0 & 1 & 0 & 0 & 0 \\
W4 & 0 & 0 & 0 & 1 & 0 & 0 & 0 & 0 & 0 & 1 & 0 & 0 \\
W5 & 0 & 0 & 0 & 0 & 1 & 1 & 0 & 0 & 0 & 1 & 0 & 1 \\
W6 & 0 & 0 & 0 & 0 & 1 & 1 & 0 & 1 & 0 & 0 & 0 & 0 \\
H1 & 0 & 0 & 0 & 0 & 0 & 0 & 1 & 0 & 0 & 0 & 0 & 0 \\
H2 & 0 & 0 & 0 & 0 & 0 & 0 & 0 & 1 & 0 & 0 & 0 & 0 \\
H3 & 0 & 0 & 0 & 0 & 0 & 0 & 0 & 0 & 1 & 0 & 1 & 0 \\
H4 & 0 & 0 & 0 & 0 & 0 & 0 & 0 & 0 & 0 & 1 & 0 & 1 \\
R1 & 0 & 0 & 0 & 0 & 0 & 0 & 0 & 0 & 1 & 0 & 1 & 1 \\
R2 & 0 & 0 & 0 & 0 & 0 & 0 & 0 & 0 & 0 & 1 & 1 & 1
\end{bmatrix}$$

(10)

### 4.2. Description of the intervention types

As shown in Fig. 5, seven preventive intervention types are to be planned. The intervention types target different objects at different locations. The intervention types with their descriptions are listed in Table 2. Information about the time intervals between interventions of the same type is also presented. Interventions of the same type can be performed with minimum and maximum time intervals $G_{\min, k}$ and $G_{\max, k}$, respectively. As previously mentioned, $G_{\min, k}$ is set to avoid unnecessary intervention while $G_{\max, k}$ to avoid unexpected failure of the object. $G_{\max, k}$ is equivalent to the mean time between failure (MTBF), which can be derived from analyses of the life expectancy of the objects.

$G_{\min, k}$ and $G_{\max, k}$ are assumed here as given since numerous methods have been developed in the past that can be used to compute their values [26,36,37,47]. In this example, intervention type $I_7$ is set as a central intervention type that starts at the first time step and occurs with a fixed time interval of three time steps. Column 7 shows the cost of executing the intervention type, $c_k$. This cost may include the cost of replacement parts, mobilizing resources, etc. Table 2 also includes a list of objects that are affected by the interventions. Interventions can directly affect multiple objects at the same time. For example, I5 is an intervention on the crossing joint of the highway H3 and railway R1. Hence, two objects are (directly) affected by this intervention type. In this case, two operators are responsible for the cost of intervention type (i.e., operators H and R). The relations between the intervention types and the objects, which are derived from Table 2, are represented by the relation matrix $R$ in Eq. (11), which indicates upon which object $i$ each intervention type $k$ intervenes.

$$\mathbf{R} = [r_{i,k}] = \begin{bmatrix}
 & I1 & I2 & I3 & I4 & I5 & I6 & I7 \\
W1 & 0 & 1 & 0 & 0 & 0 & 0 & 0 \\
W2 & 0 & 0 & 1 & 0 & 0 & 0 & 0 \\
W3 & 0 & 0 & 0 & 0 & 0 & 1 & 0 \\
W4 & 0 & 0 & 0 & 0 & 0 & 1 & 0 \\
W5 & 0 & 0 & 0 & 0 & 0 & 1 & 0 \\
W6 & 0 & 0 & 0 & 0 & 0 & 0 & 0 \\
H1 & 1 & 0 & 0 & 1 & 0 & 0 & 0 \\
H2 & 0 & 0 & 0 & 1 & 0 & 0 & 0 \\
H3 & 0 & 0 & 0 & 1 & 1 & 0 & 0 \\
H4 & 0 & 0 & 0 & 1 & 0 & 0 & 0 \\
R1 & 0 & 0 & 0 & 0 & 1 & 0 & 0 \\
R2 & 0 & 0 & 0 & 0 & 0 & 0 & 1
\end{bmatrix}$$

(11)

### 4.3. Optimization problem

The optimal intervention program is obtained through the optimization problem Eqs. (4)–(9). The number of time steps considered in this example is 18 time steps ($T = 18$). The IGA used to solve the optimizing problem has been run 500 times with different initial populations. The number of variables $T \times n_I$ is 126, where $n_I = 7$ is the number of intervention types. The parallel computing technique has been used to reduce the simulation time. The optimization problem was solved using the optimization package of Matlab® R2018b on a desktop computer with the following specifications: Windows 10, Intel Core i5–6500 CPU @3.20GHz, and installed memory (RAM) of 8 GB, and the total simulation time was 12 mins.

### 4.4. Results

Fig. 6 shows the optimal intervention program of the intervention types for a period of 18 time steps. Every row on the graph represents the intervention program of one intervention type. As can be seen, no pattern could be identified for the intervention programs of the seven intervention types. This demonstrates that finding the optimal intervention program is not intuitive especially when large networks are involved. The minimum and maximum time intervals (i.e., the constraints) between two successive interventions of the same type (see Table 2) are all satisfied. The cumulative costs of interventions $f_1$, the cumulative cost of service interruption (service unavailability cost) $f_2$, and the cumulative total cost are plotted on the same graph. It is clear that the service unavailability cost makes up most of the total cost. Therefore, it is easy to reduce the cost by better arranging the interventions, even if the arrangement does not yield the least total number of interventions. The 500 simulations (with 500 initial populations each) have resulted in 19 unique intervention programs. The cumulative curve of the total cost obtained from the 19 programs is plotted in Fig. 7. The program with the least total cost is selected as the optimal program (black curve). There is approximately a 10% difference in the final cost between the best-known solution and the most expensive sub-optimal solution. This demonstrates the robustness of the optimization algorithm. If this variability range had been much larger, exploring other algorithms would have been required. What is interesting about this graph is that the optimal solution is conditioned on the period of analysis. If the period of analysis was, for instance, 11 time steps, another solution would have been considered optimal (see dashed-line curve). To demonstrate this, another simulation for a longer period (T = 60 time steps) is performed and the results are presented in Fig. 8. It is evident in this figure that the optimal program for the first 18 time steps is different than that in Fig. 6. Therefore, it is not adequate to adopt an optimal solution obtained from, for instance, a short-term simulation for long-term planning. Hence, the period of analysis should be carefully selected. It is important to note that the total simulation time for the 500 simulations with T = 60 was 35 mins. This suggests that the running time is roughly proportional to the period of analysis (i.e., it took 12 mins with a period of analysis T = 18 time steps). This means that increasing the period of analysis does not increase the running time exponentially. This is important when a longer period of analysis is considered, for instance when replacement interventions are to be included in the planning.

**Table 2**
Description of the intervention types.

| Intervention type | Index ($k$) | Intervention on | Objects directly affected ($i$) | $G_{\min, k}$ (time step) | $G_{\max, k}$ (time step) | Cost of one intervention of type k, $c_k$ (monetary unit) $\times 10^3$ | Operator responsible (W,H,R) |
|---|---|---|---|---|---|---|---|
| I1 | 1 | Highway section H1 | 7 (H1) | 3 | 5 | 5 | H |
| I2 | 2 | Water network section W1 | 1 (W1) | 2 | 6 | 2,5 | W |
| I3 | 3 | Water network section W2 | 2 (W2) | 4 | 6 | 4 | W |
| I4 | 4 | Highway crossing J2 | 7, 8, 9, 10 (H1, H2, H3, H4) | 3 | 4 | 4,5 | H |
| I5 | 5 | Crossing of the highway section H3 and railway section R1 | 9, 11 (H3, R1) | 3 | 3 | 3 | H&R |
| I6 | 6 | Water joint J9 | 3, 4, 5 (W1, W2, W3) | 4 | 6 | 5,5 | W |
| I7 | 7 | Railway section R2 | 12 (R2) | 2 | 4 | 3 | R |

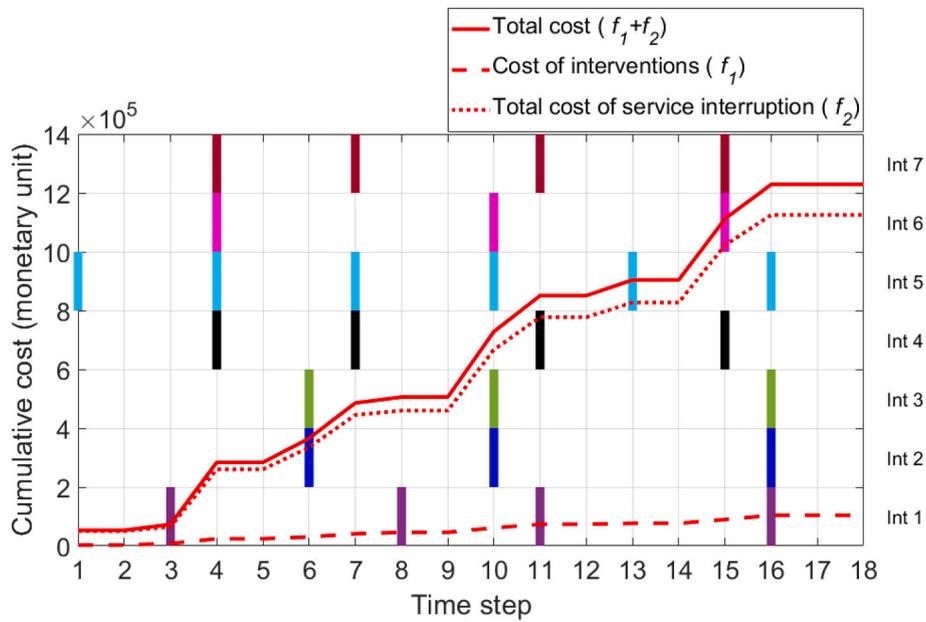

**Fig. 6.** Optimal intervention program with the corresponding cumulative incurred cost for T = 18 time steps.

To analyze the cost-benefit, the optimal intervention program is compared to another intervention program in which the number of interventions is minimum. The latter implies minimum intervention cost for the operator and thus it is usually assumed among operators [16]. In such intervention program, the time between two successive interventions of the same type is equivalent to $G_{max}$. Hereafter, this intervention program will be referred to as *individual* intervention program. The individual intervention program occurs when every operator individually plans their interventions with no regard to other operators' intervention programs, overlooking the service unavailability their interventions would cause to other networks. In real life, this is usually the case because there is indeed minimal or no communication among the infrastructure operators in this regard. Table 3 compares the cost incurred to each operator from the individual intervention program against the optimal program given by the 3C-approach.

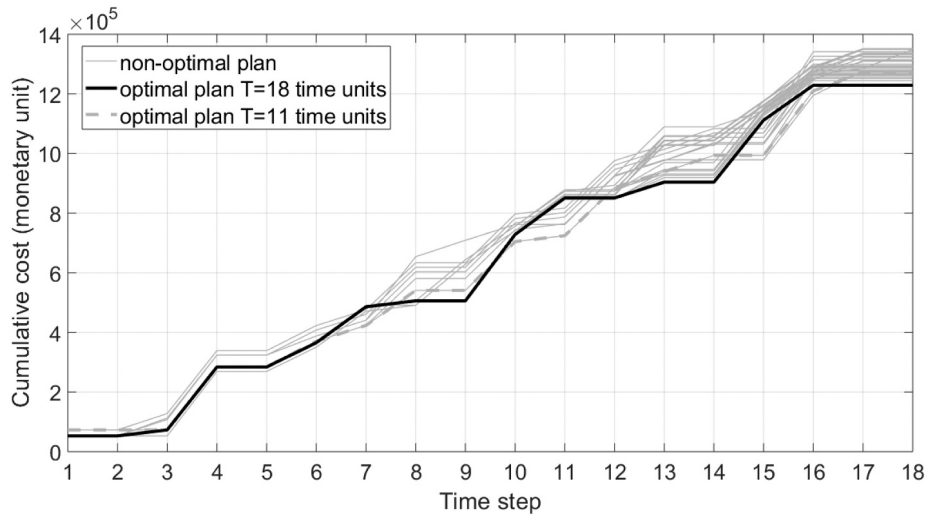

**Fig. 7.** Cost comparison between optimal and sub-optimal intervention programs.

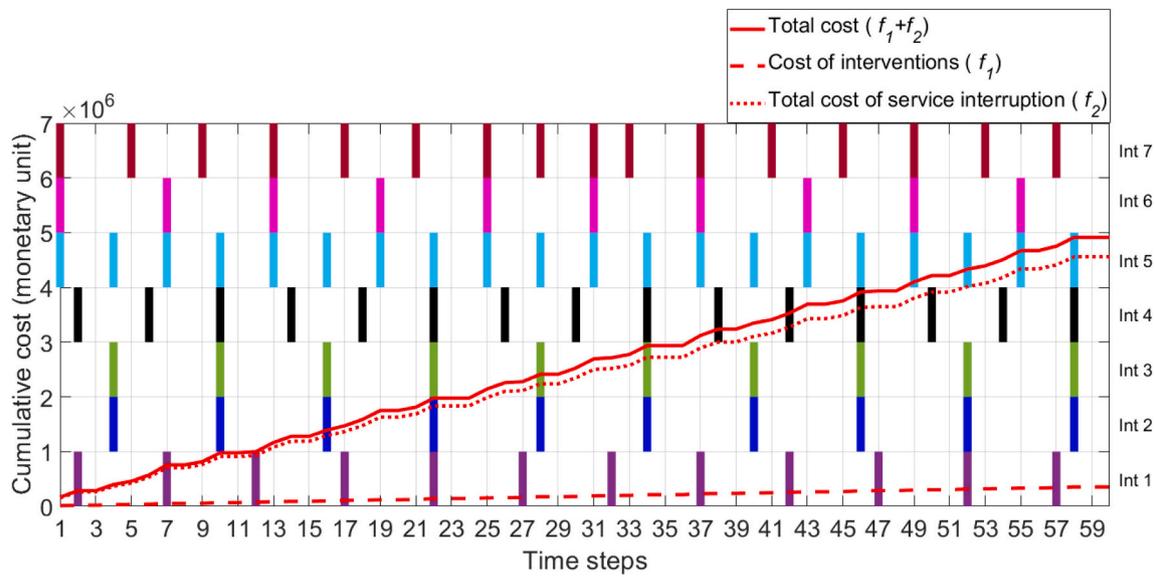

**Fig. 8.** Optimal intervention program with the corresponding cumulative incurred cost for T = 60 time steps.

The top half of the table presents the costs related to the intervention types while the bottom half presents the service unavailability costs. The ratio between the results of the optimal solution and the individual intervention program is calculated. For the intervention costs, results show an increase of 25% of the cost for the highway operator and 18% for the railway operator. These percentages are translated into 9500 and 3000 monetary units, respectively. The increase in intervention costs for the highway and railway operators is due to an increase in the number of interventions for these operators. This implies that these operators must perform more interventions than what they used to do and they are, therefore, put at a disadvantage. There are several possible ways to mitigate the additional costs paid by some of the operators. A possible approach is to adopt a sub-optimal intervention program that guarantees an equal additional cost by all operators. The disadvantage of this approach is the loss in the net benefit, which is the difference between the net benefit resulting from the optimal program and that resulting from the selected sub-optimal program. Another approach is to divide the extra costs over all operators proportionally to what they would have paid if they have had planned their own intervention programs.

The bottom half of the table shows a substantial decrease in the service unavailability costs for all operators. The total saving from the reduced service unavailability sums up to 370,000 monetary units (i.e., 25% reduction). This significant saving is the result of the optimal arrangement of interventions, which imposed additional intervention costs against two operators. The additional intervention costs can be considered as an investment towards a substantial increase in the global benefit. The total service unavailability cost saving is 30 times greater than the extra intervention costs.

## 5. Discussion and conclusions

In this paper, the integrative 3C concept, a multi-system optimization approach for infrastructure intervention planning, is presented. The proposed approach can consider the interactions between multiple infrastructure networks and multiple stakeholders (e.g., society and infrastructure operators), and can accommodate different types of interventions, such as maintenance, removal, and upgrading. The interactions within and across the infrastructure networks are modeled using a proposed interaction matrix (IM), which can account for different types of interdependencies, such as physical, geographical, and functional. What stands out in this approach is the ability to distinguish between *central* and *non-central* intervention types. Central interventions are those that occur at a pre-established time moment and are usually implemented with a fixed time interval. Non-central interventions, on the other hand, are the flexible interventions that can be clustered around other interventions, central or non-central. This distinction allows better reflecting the actual human actions during intervention planning. Finally, the proposed intervention scheduling approach is formalized using a comprehensive mathematical model.

**Table 3**
Comparison between the individual and the 3C intervention programs for T = 18 time steps.

|  | Individual approach $\sum_1^T c_k$ (monetary unit) ×10³ | Optimal solution $\sum_1^T c_{k,opt}$ (monetary unit) ×10³ | Ratio optimal/ individual (monetary unit) ×10³ | Difference (monetary unit) ×10³ |
|---|---|---|---|---|
| Operator W | 36 | 36 | 1 | 0 |
| Operator H | 37,5 | 47 | 1.25 | +9,5 |
| Operator R | 16,5 | 19,5 | 1.18 | +3 |
| **Total intervention cost** | **90** | **102,5** | **1.14** | **+12,5** |
| Operator W | 480 | 367,5 | 0.77 | −112,5 |
| Operator H | 572,5 | 435 | 0.76 | −137,5 |
| Operator R | 442,5 | 322,5 | 0.73 | −120 |
| **Total service unavailability cost** | **1495** | **1125** | **0.75** | **−370** |
| **Total Cost (intervention + service unavailability)** | **1585** | **1227,5** | **0.77** | **−357,5** |

To illustrate the applicability of the 3C concept, a numerical example of three interdependent infrastructure networks is presented. It has been shown that finding the optimal arrangement of interventions may significantly reduce the total cost, which is divided into (indirect) service unavailability cost and (direct) intervention cost. The decrease in cost, amounting to 25% in the example, is due to the reduction in the service unavailability.

The proposed optimization problem is simple and easily scalable. The time consumed by the simulation is roughly proportional to the number of variables, unlike other published algorithms where the simulation time increases exponentially by increasing the number of variables. This is a significant advantage because it allows for intervention planning of systems with many objects accounting for the interdependencies among them. Moreover, the optimal intervention program obtained in the analysis might look random at first glance; however, as the results are based on an optimization, they are better than what could be obtained by human intuition. This indicates it is a useful managerial decision support tool.

The solution obtained by the proposed optimization approach is considered conservative as compound service unavailability effects are not considered. The compound effect occurs when there is no proportionality between the level of service interruption and the degree of impact. For instance, multiple small interventions can cause a cut-off of a whole region. Nonetheless, this is not going to significantly modify the optimal point as the service unavailability impact is normally much larger than the intervention costs; thus, if a larger unavailability impact is incurred due to the compound effect, the savings obtained by applying this approach would be even larger.

The proposed method can be used for different types of intervention, such as maintenance and replacement. Replacing an object in a network can be done by introducing it as an intervention in the planning scheme. In the case that the object to be replaced is crossing with another object, a decision must be taken whether to postpone the replacement of the first object or to bring forward the replacement of the second object. This decision is reflected in the parameters $G_{min}$ and $G_{max}$ which define the time flexibility in executing an intervention. In some cases, the two objects cannot be replaced at the same time due to a large gap in their service life, then the corresponding interventions shall be executed separately, if possible. Nonetheless, it is advised that operators agree on making their intervention scheduling more flexible as it can result in large benefits.

The results of this paper encourage infrastructure managers to foster communication between each other regarding their intervention planning as this could bring significant benefits to all of them by jointly planning their interventions. Future work will be geared towards extending the 3C approach by 1) accounting for the uncertainty of interventions occurrence, 2) exploring other heuristic-based methods to improve the computational time, 3) revoking the assumption that an intervention is completed during a single time interval, which allows considering longer intervention durations, and 4) validating the 3C concept in a real-life asset management case (organizations, infrastructure data, etc.) and analyzing the optimal solutions for different stakeholders.

**Declaration of Competing Interest**

None.

**Acknowledgments**

This work is part of the research programme D-to-MII (Design to Manage Interconnected Infrastructures) with project number 439.16.823, which is financed by the Dutch Research Council (NWO), The Netherlands.